\documentclass{PoS}%
\usepackage{amsmath}
\usepackage{amsfonts}
\usepackage{amssymb}
\usepackage{graphicx}%
\providecommand{\U}[1]{\protect\rule{.1in}{.1in}}
\title{%
\protect{
\vspace{-1.2cm}
\flushright{
{\normalsize 
\begin{minipage}{4cm}
KEK Preprint 2009-28 \\%
CHIBA-EP-180
\end{minipage}
}}% %right%
\\%
\vspace{1cm}
}% 
Topological configurations of Yang-Mills field responsible %
for magnetic-monopole loops as quark confiner%
}

\ShortTitle{Topological configurations of Yang-Mills field responsible %
for magnetic-monopole loops ...}

\author{\speaker{Akihiro Shibata}\\%
Computing Research Center, High Energy Accelerator Research Organization (KEK) \& \\%
Graduate University for Advanced Studies (Sokendai), Tsukuba 305-0801, Japan\\%
E-mail: \email{akihiro.shibata@kek.jp}
}

\author{Kei-Ichi Kondo\\
Department of Physics, Graduate School of Science, Chiba University, Chiba 263-8522, Japan\\%
        E-mail: \email{kondok@faculty.chiba-u.jp}}
\author{Seikou Kato\\
Takamatsu National College of Technology, Takamatsu 761-8058, Japan\\
        E-mail: \email{kato@takamatsu-nct.ac.jp}}
\author{Shoichi Ito\\
Nagano National College of Technology, 716 Tokuma, Nagano 381-8550, Japan\\
        E-mail: \email{shoichi@ei.nagano-nct.ac.jp}}
\author{Toru Shinohara\\
Graduate School of Science, Chiba University, Chiba 263-8522, Japan\\
        E-mail: \email{Shinohara@graduate.chiba-u.jp}}

\author{Nobuyui Fukui\\
Graduate School of Science, Chiba University, Chiba 263-8522, Japan\\
        E-mail: \email{n.fukui@graduate.chiba-u.jp}}

\abstract{%
    We have given a new description of the lattice Yang-Mills theory a la Cho-Faddeev-Niemi-Shabanov, 
which has enabled us to confirm in a gauge-independent manner "Abelian"-dominance 
and magnetic-monopole dominance in the Wilson loop average,
 yielding a gauge-independent dual superconductor picture 
for quark confinement. In particular, 
we have given a new procedure (called reduction) for obtaining
 a gauge-independent magnetic monopole from a given Yang-Mills field. 
    In this talk, we demonstrate how some of known topological configurations 
in the SU(2) Yang-Mills theory such as merons 
and instantons generate closed loops of magnetic-monopole current as the quark confiner,
 both of which are characterized by the gauge-invariant topological index,
 topological charge (density) and magnetic charge (density), respectively. 
We also try to detect which type of topological configurations exist 
in the lattice data involving magnetic-monopole loops generated by Monte Carlo simulation. 
Here we apply a new geometrical algorithm based on "computational homology" 
to discriminating each closed loop from clusters of magnetic-monopole current, 
since the magnetic-monopole current on a lattice is integer valued. 

}

\FullConference{The XXVII International Symposium on Lattice Field Theory\\
		 July 26-31, 2009\\
		 Peking University, Beijing, China}

\begin{document}
\section{Introduction}

It is interesting to study the confinement mechanism of QCD. The dual
superconductivity picture is a promising mechanism for quark confinement. It
is known that the string tension of the Abelian part and the monopole part in
Yang-Mills (YM) fields reproduce the original one, so called Abelian dominance
and monopole dominance in the string tension. There is another approach that
the center vortex can explain the string tension. However, the dominances of
these objects has been observed only in the special gauges such as the maximal
Abelian (MA) gauge or the maximal center gauge, but not in other gauges.

We have given a new description of the Yang-Mills (YM) theory on a lattice,
which is expected to give an efficient framework to explain quark confinement
based on the dual superconductivity picture\ in the gauge independent manner.
In the case of the $SU(2)$ Yang-Mills theory, it is constructed as a lattice
version of the Cho-Faddeev-Niemi-Shabanov (CFNS) decomposition\cite{CFNS-C} in
a continuum theory\cite{KSM05}\cite{ref:NLCVsu2}\cite{ref:NLCVsu2-2}.  By
performing numerical simulations we have demonstrated that the gauge-invariant
magnetic monopoles can be constructed and they reproduces the string tension
\cite{ref:NLCVsu2}, and shown the infrared \textquotedblleft Abelian"
dominance in the propagator, i.e., the extracted "Abelian" propagator is
dominant in the infrared region\cite{ref:NLCVsu2-2}. We have extended the
framework to the $SU(N)$ YM theory \cite{SCGTKKS08}. In fact, we have
demonstrated the numerical simulation in the $SU(3)$ case, which has two
options corresponding to its stability subgroups. One is the maximal option
with the stability subgroup $\tilde{H}=U(1)\times U(1)$, which is the
gauge-independent reformulation of the Abelian projection represented by the
conventional MA gauge. We have demonstrated the gauge independent study of
"Abelian" dominance in the conference lattice2007 \cite{lattce2007}. The other
is the minimal one with the stability group $\tilde{H}=U(2)$, which is a new
type of formulation and derives a non-Abelian magnetic monopole
\cite{KondoNAST}. We have demonstrated the non-Abelian magnetic monopole
dominance in the conference lattice2008 \cite{lattice2008}.

In this talk, we apply this method to investigate the role of magnetic
monopole for the confinement in the gauge independent manner. In what follows,
we restrict to the $SU(2)$ case. We summarize the result of the lattice
formulation of the CFNS decomposition, which can extract the dominant degrees
of freedom that are relevant to quark confinement in the Wilson criterion in
such a way that they reproduce almost all the string tension in the linear
inter-quark potential. Combined with non-Abelian Stokes's theorem, the
magnetic monopoles can be extracted from the decomposed relevant part of
YM\ field in the gauge invariant manner. The monopole dominance suggests that
the magnetic monopole plays a central role in quark confinement. So, it is
interesting to investigate the magnetic monopole as a quark confiner. We
demonstrate the analysis for some known topological configurations as well as
lattice data, where we can extract magnetic monopole loops directly from YM
configuration, which are the gauge invariant physical objects.

\section{New variables on a lattice}

We summarize the new description of the $SU(2)$YM theory on a
lattice\cite{ref:NLCVsu2}\cite{ref:NLCVsu2-2} (See also
\cite{kato:lattice2009}): The YM field $\mathbb{A}_{x^{\prime},\mu}%
\mathbf{\ }$is represented as a link variable%
\begin{equation}
U_{x,%
%TCIMACRO{\U{3bc} }%
%BeginExpansion
\mu
%EndExpansion
}=\exp\left(  -ig\int_{x}^{x+\hat{\mu}\epsilon}dx^{\mu}\mathbf{A}_{\mu
}(x)\right)  =\exp\left(  -ig\epsilon\mathbb{A}_{x^{\prime},\mu}\right)  ,
\end{equation}
which is supposed to be decomposed into the product of new variables,
$X_{x,\mu}$, $V_{x,\mu}\in SU(2)$,%
\begin{subequations}
\begin{align}
U_{x,\mu} &  =X_{x,\mu}V_{x,\mu}\label{decomp}\\
V_{x,%
%TCIMACRO{\U{3bc} }%
%BeginExpansion
\mu
%EndExpansion
} &  =\exp\left(  -ig\epsilon\mathbb{V}_{x^{\prime},\mu}\right)
,\label{Var:V}\\
X_{x,%
%TCIMACRO{\U{3bc} }%
%BeginExpansion
\mu
%EndExpansion
} &  =\exp\left(  -ig\epsilon\mathbb{X}_{x,\mu}\right)  ,\label{var:X}%
\end{align}
such that the decomposed variables are transformed by a full gauge
transformation $\Omega_{x}\in SU(2)$:%
\end{subequations}
\begin{subequations}
\begin{align}
U_{x,%
%TCIMACRO{\U{3bc} }%
%BeginExpansion
\mu
%EndExpansion
} &  \rightarrow U_{x,%
%TCIMACRO{\U{3bc} }%
%BeginExpansion
\mu
%EndExpansion
}^{\prime}=\Omega_{x}U_{x,%
%TCIMACRO{\U{3bc} }%
%BeginExpansion
\mu
%EndExpansion
}\Omega_{x+\mu}^{\dag},\label{eq:GTU}\\
V_{x,%
%TCIMACRO{\U{3bc} }%
%BeginExpansion
\mu
%EndExpansion
} &  \rightarrow V_{x,%
%TCIMACRO{\U{3bc} }%
%BeginExpansion
\mu
%EndExpansion
}^{\prime}=\Omega_{x}V_{x,%
%TCIMACRO{\U{3bc} }%
%BeginExpansion
\mu
%EndExpansion
}\Omega_{x+\mu}^{\dag},\qquad X_{x,%
%TCIMACRO{\U{3bc} }%
%BeginExpansion
\mu
%EndExpansion
}\rightarrow X_{x,%
%TCIMACRO{\U{3bc} }%
%BeginExpansion
\mu
%EndExpansion
}^{\prime}=\Omega_{x}X_{x,%
%TCIMACRO{\U{3bc} }%
%BeginExpansion
\mu
%EndExpansion
}\Omega_{x}^{\dag},\label{eq:GTV}%
\end{align}
where $V_{x,\mu}$ is defined on a link $\left\langle x,x+\epsilon
\mu\right\rangle $ like $U_{x,%
%TCIMACRO{\U{3bc} }%
%BeginExpansion
\mu
%EndExpansion
}$, and $X_{x,\mu}$ on a site. So, the corresponding (Lie algebra-valued)
gauge filed is evaluated at the midpoint, $x^{\prime}=x+\epsilon\mu/2,$ for
$U_{x,\mu}$ and $V_{x,%
%TCIMACRO{\U{3bc} }%
%BeginExpansion
\mu
%EndExpansion
},$ and at the site $x$ for $X_{x,\mu}.$To define the decomposition, a color
field, $\mathbf{n}_{x}=n_{x}^{k}\sigma^{k}/2$ is introduced as a site
variable, where $\sigma^{k}$ is the Pauli matrix and $n_{x}^{k}$ $(k=1,2,3)$
is a unit vector. The color field is transformed adjointly by an independent
gauge transformation $\Theta_{x}$ $\in SU(2)$ as $\,\mathbf{n}_{x}\rightarrow$
$\Theta_{x}\mathbf{n}_{x}\Theta_{x}^{\dag}$. The decomposition is determined
by solving the defining equations:
\end{subequations}
\begin{subequations}
\label{eq:define}%
\begin{align}
&  D_{\mu}^{\epsilon}[V]\mathbf{n}_{x}:=\mathbf{n}_{x}V_{x,%
%TCIMACRO{\U{3bc} }%
%BeginExpansion
\mu
%EndExpansion
}-V_{x,%
%TCIMACRO{\U{3bc} }%
%BeginExpansion
\mu
%EndExpansion
}\mathbf{n}_{x+\mu}=0,\\
&  \mathrm{tr}(\mathbf{n}_{x}X_{x,%
%TCIMACRO{\U{3bc} }%
%BeginExpansion
\mu
%EndExpansion
})=0,
\end{align}
and the solution is obtained in terms of \ $U_{x,%
%TCIMACRO{\U{3bc} }%
%BeginExpansion
\mu
%EndExpansion
}$ and $\mathbf{n}_{x}$
\end{subequations}
\begin{subequations}
\begin{align}
V_{x,%
%TCIMACRO{\U{3bc} }%
%BeginExpansion
\mu
%EndExpansion
} &  =\tilde{V}_{x,%
%TCIMACRO{\U{3bc} }%
%BeginExpansion
\mu
%EndExpansion
}/\sqrt{\frac{1}{2}\mathrm{tr}\left(  \tilde{V}_{x,%
%TCIMACRO{\U{3bc} }%
%BeginExpansion
\mu
%EndExpansion
}\tilde{V}_{x,%
%TCIMACRO{\U{3bc} }%
%BeginExpansion
\mu
%EndExpansion
}^{\dag}\right)  },\qquad\tilde{V}_{x,%
%TCIMACRO{\U{3bc} }%
%BeginExpansion
\mu
%EndExpansion
}=U_{x,%
%TCIMACRO{\U{3bc} }%
%BeginExpansion
\mu
%EndExpansion
}+4\mathbf{n}_{x}U_{x,%
%TCIMACRO{\U{3bc} }%
%BeginExpansion
\mu
%EndExpansion
}\mathbf{n}_{x+\mu},\\
X_{x,\mu} &  =U_{x,%
%TCIMACRO{\U{3bc} }%
%BeginExpansion
\mu
%EndExpansion
}V_{x,%
%TCIMACRO{\U{3bc} }%
%BeginExpansion
\mu
%EndExpansion
}^{\dag}.
\end{align}
The reduction condition must be introduced in order that the theory written in
terms of new variables is equipollent to the original YM theory, i.e., the
symmetry extended by introducing the color field, $SU(2)_{\Omega}\times\left[
SU(2)/U(1)\right]  _{\Theta},$ must be reduced to the same symmetry as the
original YM theory, $SU(2)_{\Omega=\Theta}.$ The reduction condition is given
by the minimization of the functional, which is invariant under the gauge
transformation, $SU(2)_{\Omega=\Theta}$. Here, we use the following
functional:%
\end{subequations}
\begin{equation}
F_{\text{Red.}}=\sum\nolimits_{x,\mu}\mathrm{tr}\left(  \left(  D_{\mu
}^{\epsilon}[U]\mathbf{n}_{x}\right)  \left(  D_{\mu}^{\epsilon}%
[U]\mathbf{n}_{x}\right)  ^{\dag}\right)  /\mathrm{tr}\left(  \mathbf{1}%
\right)  .
\end{equation}
Note that the stationary condition with respect to the color field, $\partial
F_{\text{Red.}}/\partial n_{x}^{k}=0,$ gives the differential form
corresponding to the continuum theory, and it determines the color field for a
given YM field. The algorithm solving the reduction condition is given by
Ref.\cite{kato:lattice2009}.

\section{Gauge independent magnetic monopole}

The Wilson loop operator on a lattice is given by the path ordered product of
link variables,%
\begin{equation}
W_{C}[U]:=\mathrm{tr}\left[  P\prod\nolimits_{<x,x+\mu>\in C}U_{x,\mu}\right]
/\mathrm{tr}(\mathbf{1})=\mathrm{tr}\left[  P\exp\left(  -ig%
%TCIMACRO{\doint _{C}}%
%BeginExpansion
{\displaystyle\oint_{C}}
%EndExpansion
dx^{\mu}\mathbf{A}_{\mu}(x)\right)  \right]  /\mathrm{tr}(\mathbf{1}),
\end{equation}
where the path $C$ is defined along the relevant links, and we have used the
definition of the link variable, $U_{x,\mu},$ in the 2nd equality. Following
the paper \cite{KondoNAST}, the Wilson loop operator in the fundamental
representation in the continuum theory is rewritten into
\begin{align}
W_{C}[\mathbf{A}] &  :=\mathrm{tr}\left[  P\exp\left(  -ig%
%TCIMACRO{\doint _{C}}%
%BeginExpansion
{\displaystyle\oint_{C}}
%EndExpansion
dx^{\mu}\mathbf{A}_{\mu}(x)\right)  \right]  /\mathrm{tr}(\mathbf{1}%
)\nonumber\\
&  =\int d\mu\lbrack\xi]_{C}\exp\left\{  -ig%
%TCIMACRO{\doint _{C}}%
%BeginExpansion
{\displaystyle\oint_{C}}
%EndExpansion
dx^{\mu}\mathbf{V}_{\mu}(x)\right\}  \nonumber\\
&  =\int d\mu\lbrack\xi]_{\Sigma}\exp\left\{  -ig\int_{\Sigma:\partial
\Sigma=C}dS^{\mu\nu}\mathcal{F}_{\mu\nu}[\mathbf{V}]\right\}  \label{eq:Wc[A]}%
\end{align}
where $\mathbf{V}_{\mu}(x)$ is given by
\begin{equation}
\mathbf{V}_{\mu}(x)=\mathrm{tr}(\mathbf{A}_{\mu}(x)\mathbf{n}(x))\mathbf{n}%
(x)+\frac{1}{ig}\left[  \partial_{\mu}\mathbf{n}(x),\mathbf{n}(x)\right]  .
\end{equation}
Following the paper \cite{KondoShibata}, it is turn out that the defining
equation eq(\ref{eq:define}) gives the decomposition which reproduces
"Abelian" ($V$) dominance for the Wilson loop operator on a lattice even with
a finite lattice spacing $\epsilon$ ;
\begin{equation}
W_{C}[U]\cong\text{\textrm{const.}}W_{C}[V],\qquad W_{C}[V]:=\mathrm{tr}%
\left[  P\prod\nolimits_{<x,x+\mu>\in C}V_{x,\mu}\right]  /\mathrm{tr}%
(\mathbf{1})
\end{equation}
and we can identify the $\mathbb{V}_{x^{\prime},\mu}$ in eq(\ref{Var:V})
with$\mathbf{V}_{\mu}(x^{\prime})$ at the midpoint of the link. The field
strength of $\mathbf{V}$ is given by
\begin{equation}
V_{x,\alpha}V_{x+\alpha,\beta}V_{x+\beta,\alpha}^{\dag}V_{x,\beta}^{\dag}%
=\exp\left(  -ig\epsilon\mathrm{tr}\left(  \mathcal{F}_{x^{\prime\prime}%
,\mu\nu}[\mathbb{V}]\mathbf{n}_{x^{\prime}}\right)  \mathbf{n}_{x^{\prime}%
}\right)  .
\end{equation}
By using the Hodge decomposition, eq(\ref{eq:Wc[A]}) is further rewritten
into
\begin{equation}
W_{C}[\mathbf{A}]=\int d\mu\lbrack\xi]_{\Sigma}\exp\left\{  -ig(K,\Xi_{\Sigma
})-ig(J,N_{\Sigma})\right\}  ,\label{eq:NAST0}%
\end{equation}
where we have used $K:=\delta{}^{\ast}F$, $J:=\delta F$ for $F$ being the
field strength 2-form $\mathcal{F}_{\mu\nu}$, and $\ \Xi_{\Sigma}:=\delta
{}^{\ast}S_{\Sigma}\Delta^{-1}$ and $N_{\Sigma}:=\delta\Theta_{\Sigma}%
\Delta^{-1}$ with the four-dimensional Laplacian, $\Delta=d\delta+\delta d$.
Here $S_{\Sigma}$ is the vorticity tensor defined by $S_{\Sigma}^{\mu\nu}%
=\int_{\Sigma}dS^{\mu\nu}(X(\sigma))\delta(x-X(\sigma))$ on the surface
$\Sigma:\partial\Sigma=C$ supported by the Wilson loop $C$. \ Therefore, the
monopole dominance for the lattice Wilson loop is given by%
\begin{subequations}
\begin{equation}
\left\langle W_{C}[U]\right\rangle \cong\left\langle W_{C}[V]\right\rangle
=\left\langle \exp\left(  -ig\epsilon\sum\nolimits_{x,\mu}k_{x,\mu}\Xi_{x,\mu
}\right)  \right\rangle ,\label{eq:WcMono}%
\end{equation}
where the lattice magnetic monopole current $K_{x,\mu}$ and $\Xi_{x,\mu}$ are
given by%
\end{subequations}
\begin{align}
\Xi_{x,\mu} &  =%
%TCIMACRO{\dsum \nolimits_{s}}%
%BeginExpansion
{\displaystyle\sum\nolimits_{s}}
%EndExpansion
\frac{1}{2}\epsilon_{\mu\alpha\beta\gamma}\partial_{\alpha}^{s}\Delta
^{-1}(x-s)S_{\beta\gamma}^{J}(s+\mu),\text{ }\partial_{\alpha}S_{\alpha\beta
}^{J}=J_{\beta}\text{ }\in C\\
k_{x,\mu} &  =2\pi\bar{k}_{x,\mu}=\frac{1}{2\epsilon^{2}}\epsilon_{\mu
\lambda\alpha\beta}\partial_{\lambda}\Theta_{x,\alpha\beta}[V],\quad
\Theta_{x,a\beta}[V]:=\arg\mathrm{tr}\left(  (\mathbf{1+n}_{x})V_{x,\alpha
}V_{x+\alpha,\beta}V_{x+\beta,\alpha}^{\dag}V_{x,\beta}^{\dag}\right)  ,
\end{align}
where $\partial_{\lambda}$ denotes the forward difference (lattice derivative)
in the $\hat{\lambda}$ direction: $\epsilon\partial_{\lambda}%
f(x):=f(x+\epsilon\hat{\lambda})-f(x)$. It should be noticed that the magnetic
monopole current $k_{x,\mu}$ is gauge invariant, as can be seen from the
transformation law of the new variables.

\section{Analysis of the magnetic monopoles as quark confiners}

We have demonstrated the "Abelian" dominance and the magnetic-monopole
dominance in the gauge invariant way\cite{kato:lattice2009}. The static
potential calculated from the decomposed variable $V$ \ is reproduced by one
from the Yang-Mills field, i.e., the string tension obtained by $V$ field
$94\%$ of one from YM field. The string tension calculated only from magnetic
monopoles reproduces $93\%$ of it. This shows that \ the proposed new
description of YM theory extract the degrees of freedom relevant for the quark
confinement from YM field, and the magnetic monopole which is the decomposed
part of $V_{x,\mu}$, plays a central role for the quark confinement. So we
investigate the magnetic monopoles as quark confiners.

The magnetic monopole current is a integer-valued, $\bar{k}_{x,\mu}\in
\{0,\pm1,\pm2\},$ and can be identified with the link variables on the dual
lattice $\left\langle \tilde{x}-\epsilon\hat{\mu},\tilde{x}\right\rangle $
with $\tilde{x}=(x_{1}+\epsilon/2$, $x_{2}+\epsilon/2$, $x_{3}+\epsilon/2$,
$x_{4}+\epsilon/2).$ So non-zero monopole currents can be identified with
directional edges which construct a directional graph. The current
conservation,%
\[
\epsilon\partial_{\mu}k_{x,\mu}=2\pi\sum\nolimits_{\mu}\left(  \bar{k}%
_{x+\mu,\mu}-\bar{k}_{x,\mu}\right)  =0,
\]
implies that there is no source and no sink. When we identify a monopole
current of charge $\bar{k}_{x,\mu}=\pm2$ \ with a double edges of a single
charge, each vertex (site of dual lattice) has the same number of incoming and
outgoing edges. Therefore, the monopole currents construct the loops. Here,
some loops are connected each other at dual lattice sites, or share links of
the dual lattice carrying the double monopole charge.\ Thus the analysis of
the monopole configuration is converted to a geometrical problem. In the
magnetic part of the Wilson loop operator eq(\ref{eq:NAST0}) or
eq(\ref{eq:WcMono}), the term $(K,\Xi_{\Sigma})$ represents the 4-dimentional
version of the Gauss's linking number formula, that is, the linking number
between a single loop $K\,$\ and a two-dimensional surface $\Sigma$ whose
boundary is the Wilson loop $C$. The conjecture on translation the
contribution of center vortex and of magnetic monopole loops to the Wilson
loops in Ref. \cite{KondoNAST} can be checked by using the lattice data.

\begin{figure}[ptb]
\begin{center}
\includegraphics[
height=3.5cm
]{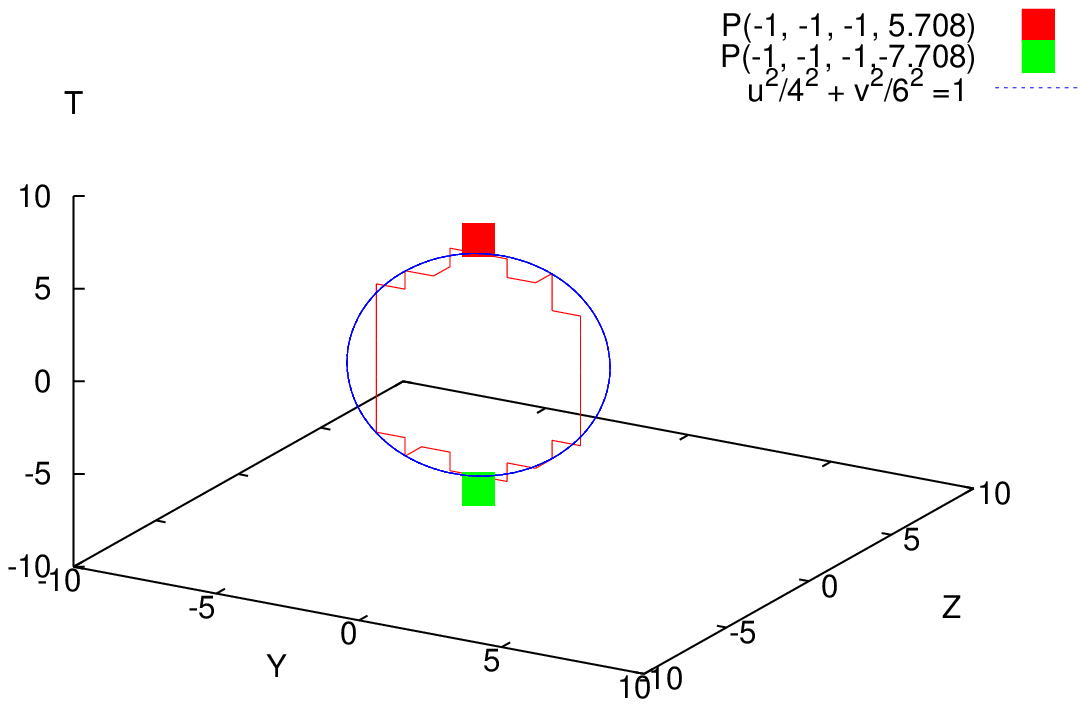} \ \includegraphics[
height=3.5cm
]{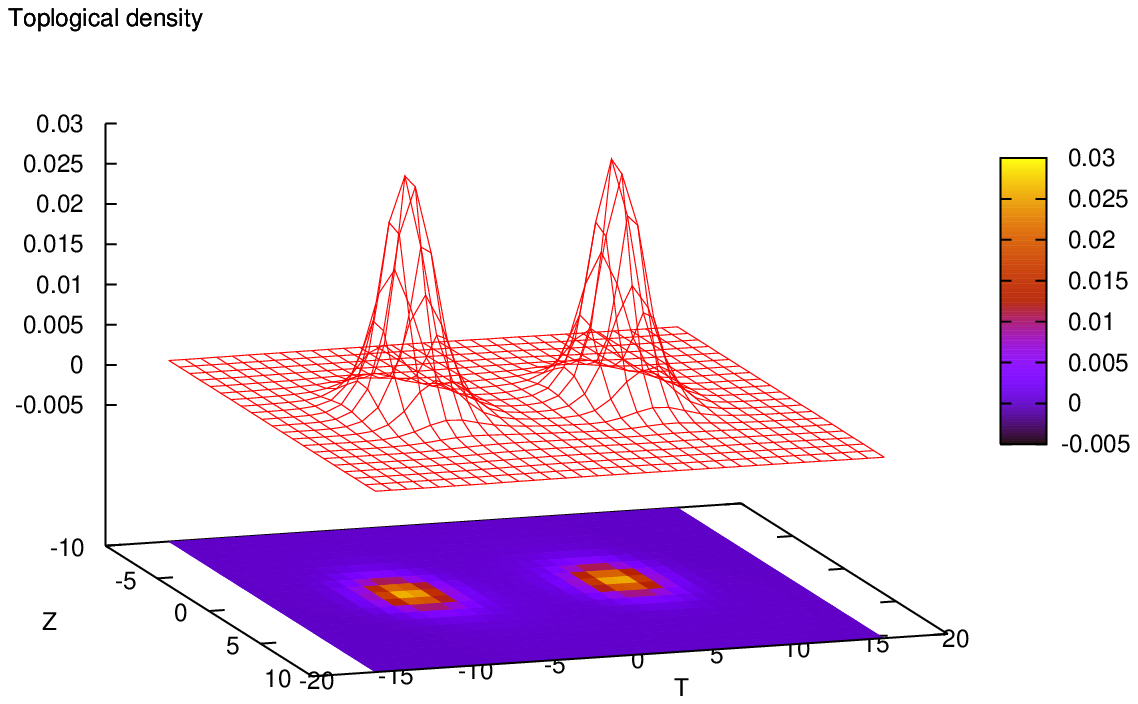}
\end{center}
\caption{(Left panel) The plot of a magnetic-monopole loop generated by a pair
of (smeared) merons in 4-dimensional Euclidean space. The 3-dimensinal plot is
obtained by projecting the 4-dimensional dual lattice space to the
3-dimensional one, i.e., ($x,y,z,t)$ $\rightarrow$ ($y,z,t$). The positions of
two meron sources are described by solid boxes, and the monopole loop by red
solid line. (Right panel) The plot of the topological charge density for $z-t$
plane  (slice of $x=y=0$). Two peaks of the topological charge density are
located at the positions of two merons.}%
\label{fig:meron}%
\end{figure}

Now, let us apply this method to topological configurations of Yang-Mills
field, which has been considered to play the central role for confinement. The
topological solution can be translated to the lattice variable by using the
definition of the link variable:%
\[
U_{x,\mu}=P\exp\left(  -ig\int_{x}^{x+\hat{\mu}\epsilon}dx^{\mu}%
\mathbf{A}_{\mu}(x)\right)  \simeq P\prod\nolimits_{n=1}^{N}\exp\left(
-ig\frac{\epsilon}{N}\mathbf{A}_{\mu}(x+(n-1/2)\frac{\epsilon}{N})\right)  ,
\]
where the integral along the link $\left\langle x,x+\hat{\mu}\epsilon
\right\rangle $ is calculated by the path ordered product of the exponentiated
gauge potential. Here, we demonstrate the two-merons case. Figure
\ref{fig:meron} shows the detected magnetic monopole. The numerical analysis
on the lattice reproduces the result of analytical study in
Ref.\cite{ref:twomeron}, even though the analysis is done by using the finite
volume lattice. It is interesting to investigate other topological
configurations such as two-instantons and calorons. We find a monopole loop
for the JNR type of two-instantons solution, while no monopole loop for
one-instanton and two-instantons of 't Hooft type. The detail analysis of
two-instantons solutions will appear in the subsequent
paper\cite{ref:mloop-twoInstanton}.

\begin{figure}[ptb]
\begin{center}
\includegraphics[
height=3.5cm
]
{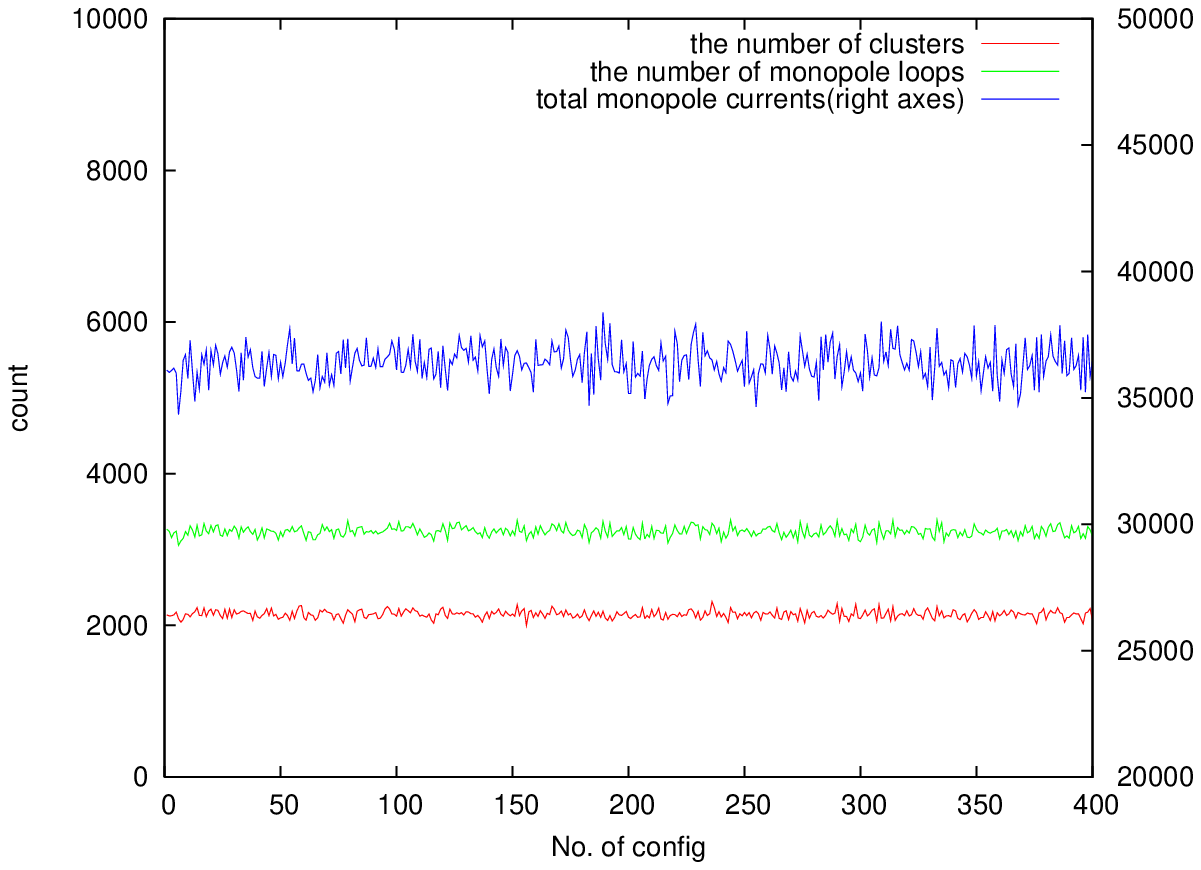} \ \ \ \includegraphics[
height=4cm
]
{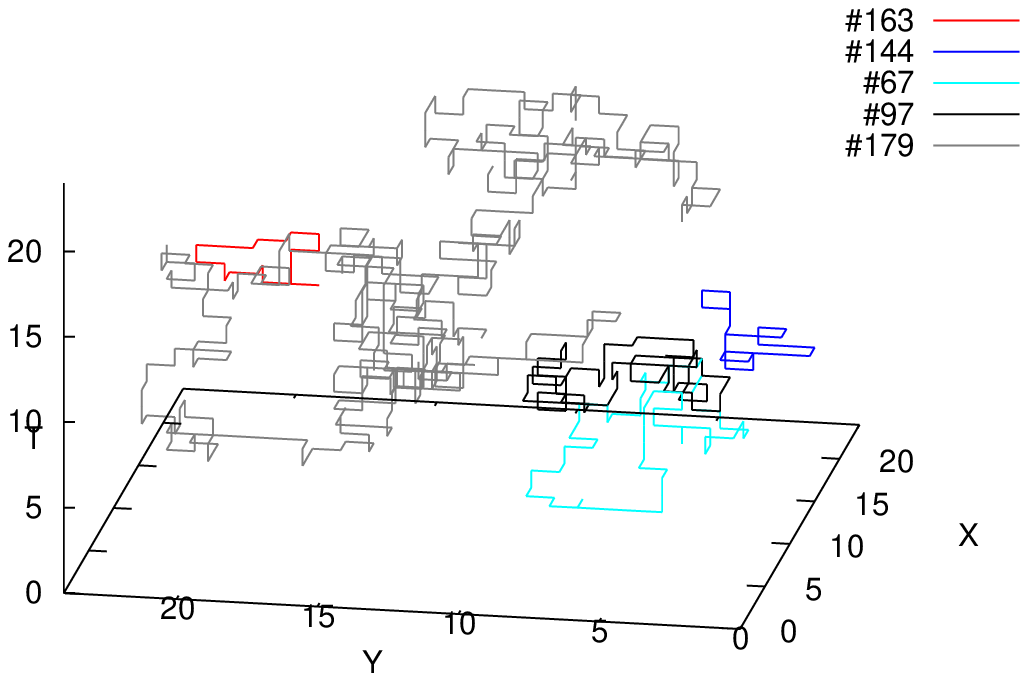}
\end{center}
\caption{(Left panel) The analysis of magnetic monoples for 400 configulations
in $24^{4}$ lattice with the parameter $\beta=2.4.$ The blue line shows the
number of the non-zero magnetic monoole currents (right virtical axes). The
red line an dgreen line show the number of clusters of connected loops (the
Betti number of dimension zero) and the number of loops (the Betti number of
dimension one) for each configuration, respectively (see left virtical axes).
(Right panel) The 3-dimensional polt of detected magbetc monopole loops, where
the graph in 4dimensional Euclid space is projected to the 3-dimensional
space, i.e., $(x,y,z,t)$ $\rightarrow(x,y,z).$ }%
\label{fig:cluster}%
\end{figure}

Then, we return to the analysis of lattice data. It is very hard to manipulate
monopole configurations directly,\ since they contain more than 35000 non-zero
monopole currents (see the left panel of Figure \ref{fig:cluster}). Therefore,
we introduce the algebraic algorithm for topology. The CHomP homology
software,\ provided by the computational homology project\cite{ref:chomp},
computes topological invariants called the Betti numbers of a collection and
their generators in the algebraic way. The Betti numbers are part of the
information contained in the homology groups of a topological space, which
intuitively measure the number of connected components, the number of holes,
and the number of enclosed cavities in low dimensions. In our case, the
generators of the dimension-one homology group correspond to magnetic monopole loops.

\begin{figure}[ptb]
\begin{center}
\includegraphics[
height=3.5cm
]
{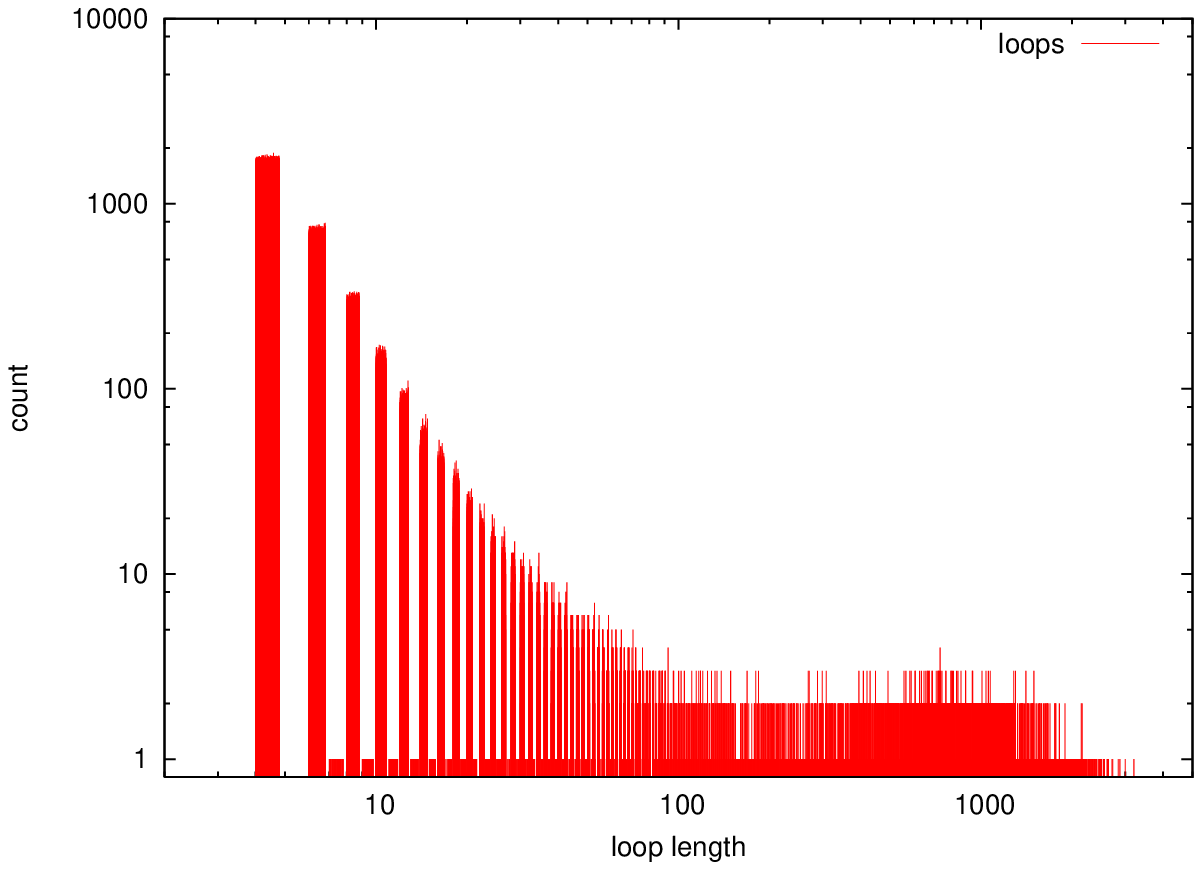} \ \ \ \ \ \ \ \includegraphics[
height=3.5cm
]
{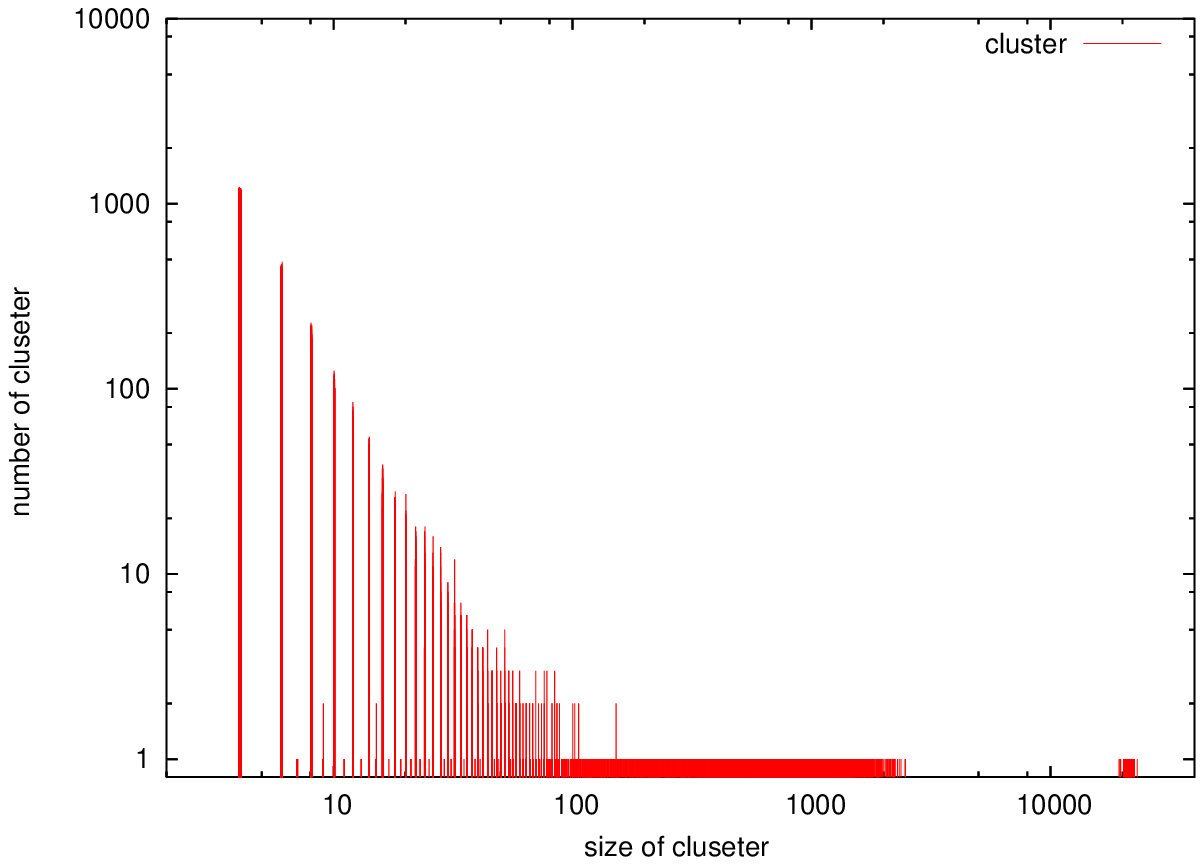}
\end{center}
\caption{(Left panel) The histgram for the length of monopole loops, i.e., for
each configuration the number of monopole loops with length $n$ is counted,
and \ is polted as a truss of poles with the same length$.$ (Right panel) The
histogram of the cluster size. }%
\label{fig:anatomy1}%
\end{figure}

We apply the method to the lattice data of $24^{4}$ lattice with periodic
boundary condition whose configurations are generated by using the standard
Wilson action with the parameter $\beta=2.4.$ The left panel of Figure
\ref{fig:cluster} shows data of detected magnetic monopole currents for 400
configurations. The blue line shows the number of non-zero charge currents,
i.e., non-zero magnetic monopole currents share about 3\% of links. The red
line and green line show the number of clusters and the number of loops for
each configurations, respectively. The right panel of Figure \ref{fig:cluster}
shows an example of detected magnetic monopoles, which are plotted in the
3-dimensional space projected from the 4-dimendional Euclidian space. Figure
\ref{fig:anatomy1} shows detail of the magnetic monopole configurations. Then,
we are ready to investigate the magnetic monopole contribution to the static
potential by using extracted monopole loops.

\section{Summary and discussions}

We have given a new description of Yang-Mills theory on a lattice, which gives
the decomposition of the Yang-Mills field for extracting the relevant degrees
of freedom for quark confinement in the gauge independent manner based on the
dual superconductivity picture. The extracted magnetic monopoles explain the
string tension and they should play the central role for quark confinement. It
is interesting to investigate the magnetic monopole as a quark confiner. The
implication of the quark confinement by the configurations topological charge
can be convert to the geometrical relations of the topological charge density
and distribution of the magnetic monopole loops, since the magnetic monopole
and topological charge density are given as a gauge invariant object (physical
object). We have applied the method to the topological configurations and the
lattice data, and extracted the magnetic-monopole loops directly. Using these
magnetic-monopole configurations, we investigate the relations among the
topological charge density, magnetic monopoles and static potential. It is
also interesting to investigate the implication of the magnetic monopole loops
for the confinement and deconfinement phase transition.

%

%TCIMACRO{\TeXButton{acknowledgement}{\section*{Acknowledgement}}}%
%BeginExpansion
\section*{Acknowledgement}%
%EndExpansion

This work is supported by the Large Scale Simulation Program No.08-16 (FY2008)
and No.09-15 (FY2009) of High Energy Accelerator Research Organization (KEK).
The numerical simulations have been done in part on a supercomputer (NEC SX-8)
at Research Center for Nuclear Physics (RCNP), Osaka University. This work is
financially supported by Grant-in-Aid for Scientific Research (C) 21540256
from Japan Society for the Promotion of Science (JSPS).

%

%TCIMACRO{\TeXButton{REFERENCES}{}}%
%BeginExpansion
%EndExpansion

\end{document}